\documentstyle[aps,pre,multicol,epsf]{revtex}

\begin{document}
\draft

\title{Fractionation of polydisperse systems: multi-phase coexistence}
\author{R. M. L. Evans\cite{email}}
\address{Department of Physics and Astronomy, The University of
Edinburgh, Edinburgh EH9 3JZ, Scotland}

\date{12th October, 1998}
\maketitle

\begin{abstract}
The width of the distribution of species in a polydisperse system is
employed in a small-variable expansion, to obtain a well-controlled and
compact scheme by which to calculate phase equilibria in multi-phase
systems. General and universal relations are derived, which determine the
partitioning of the fluid components among the phases. The analysis applies
to mixtures of arbitrarily many slightly-polydisperse components. An
explicit solution is approximated for hard spheres.
\end{abstract}


\begin{multicols}{2}
\narrowtext

It is vital to gain an understanding of polydispersity, due to its
ubiquity in both synthetic and biological complex fluids. A polydisperse
substance is a mixture of infinitely many components, and can, in general,
separate into arbitrary numbers of coexisting phases.  These properties
typically engender great mathematical complexity, and have been a
stumbling block to the concise formulation of polydisperse thermodynamics.
Experimental \cite{Koningsveld77,Bibette91,Clarke95} and simulational
\cite{Stapleton90,Bolhuis96} studies of polydisperse polymeric fluids
and colloidal suspensions have catalogued diverse behavior and
intricate phase diagrams. Until recently, theoretical treatments of
polydispersity relied on uncontrolled approximations \cite{Salgi93} or
idealised models \cite{Gualtieri82,Briano84}, while generic schemes and
fundamental understanding remained elusive.

The phase behavior of {\em pure} ({\it i.e}.~monodisperse) systems is
(in principle at least) relatively straightforward to analyse. The
standard method, formulated last century \cite{Reif}, involves
integrating (by various approximate methods) the Boltzmann factor over
all configurations to construct
the Helmholtz free energy as a function of temperature, density and
volume. From this, the densities of coexisting phases can be calculated
and the phase diagram deduced. One source of difficulty in analysing the
phase equilibria of polydisperse systems is that the density alone does
not fully characterize a phase. Instead, we wish to know the entire
{\em composition} of each phase. That requires the evaluation
of an infinite set of variables.

Two systematic schemes were developed recently to solve the polydisperse
phase equilibria problem. The powerful `annealed moments' method of
Sollich and Cates \cite{Sollich98} and Warren \cite{Warren98} applies to a
large subset of model systems, and will not be further discussed here. The
second scheme, which applies to real systems, was developed by the
present author \cite{Evans98}. It uses the {\em width} of
the distribution of species as a small expansion parameter, and is
therefore valid for {\em slightly} polydisperse systems. In other words,
the scheme is applicable whenever the polydisperse property ({\it
e.g}.~the {\em size} or {\em charge} of a particle) varies only a little
throughout the system. (N.B. Here, `particle' is used to denote the
polydisperse fluid elements, which could be polymer molecules, colloidal
latices, {\it etc}.) The method was used \cite{Evans98} to find
the complete distributions of species in two coexisting phases of
any slightly polydisperse system, resulting in a universal law of
fractionation. In this paper, the method is applied to
coexistence between arbitrary numbers of phases --- a situation of
importance to many polydisperse substances \cite{Fredrickson98}.

A slightly polydisperse system (one with a narrow distribution of
species) is in principle very different from a truly pure one, which has
no mixing entropy and whose distribution is a Dirac delta function.
Nevertheless, one would expect the physical properties of the two
systems to be very similar. That similarity motivates this study, since
the pure system is vastly simpler to analyse than its polydisperse
counterpart. To exploit that simplicity, a formalism is required which
treats mono- and poly-disperse systems on an equal footing. Such a formalism
is now derived (extending the method for two-phase equilibrium in
Ref.~\cite{Evans98}).

Let us first define a number $\varepsilon_{i}$ to characterize each of
the $N$ particles in the system (with $i=1,...,N$) \cite{comment}. For
size polydispersity, this is the fractional difference
$\varepsilon_{i}\equiv (R_{i}-R_{0})/R_{0}$ of the particle's radius
$R_{i}$ from some reference length $R_{0}$ (with the obvious generalization
to charge polydispersity {\it etc.}). Henceforth,
$\varepsilon$ shall be referred to as the size parameter, for
definiteness. The population of species in the system is
characterized by a continuous distribution $f(\varepsilon)$, which is
unnormalized so
\begin{displaymath}
  \int_{-\infty}^{\infty} f(\varepsilon)\,d\varepsilon = N.
\end{displaymath}
In general, the free energy $F$ of a polydisperse system is a
complicated functional of $f(\varepsilon)$. It will be expressed in
units of $k_{B}T$ where $k_{B}$ is Boltzmann's constant and $T$ is
temperature. For a polydisperse ideal gas, the free energy $F^{\rm id}$ is
easily shown \cite{Gualtieri82,Warren98} to be, per unit volume,
\begin{equation}
\label{ideal}
  \frac{F^{\rm id}}{V} = \int d\varepsilon\, 
  \frac{f(\varepsilon)}{V}\, \left[ \ln \frac{f(\varepsilon)}{V} - 1 \right]
\end{equation}
which is the usual ideal gas free energy, summed over all species. As this
expression contains the mixing entropy, it is useful to write the
free energy of a non-ideal system as
\begin{equation}
\label{split}
  F\equiv F^{\rm id}+F^{\rm ex}.
\end{equation}
Here $F^{\rm ex}$ is the `excess' part of the free energy (over and above
the ideal part), deriving from interactions.

Let us consider a system whose `initial' population (before phase
separation) is known. This will be called the `parent' distribution
$f_{P}(\varepsilon)$. In a system where this parent is partitioned into
${\cal M}$ coexisting phases, we wish to determine the distribution
$f(\varepsilon)_{{\cal A}}$ in each phase ${\cal A}=1,...,{\cal M}$. By
conservation of matter,
\begin{equation}
\label{conserve}
  \sum_{{\cal A}=1}^{{\cal M}} f(\varepsilon)_{{\cal A}} = f_{P}(\varepsilon).
\end{equation}
At equilibrium, the chemical potential is equal in all coexisting
phases. This statement applies for each species of particles, so the
equation
\begin{equation}
\label{equilib}
  \mu(\varepsilon)_{{\cal A}} =
  \mu(\varepsilon)_{{\cal B}} \hspace{6mm} \forall \;\; \varepsilon
\end{equation}
represents an uncountable infinity of thermodynamic constraints for any
pair of phases ${\cal A}$ and ${\cal B}$. Since there is a continuum of
species, the chemical potential is a {\em functional} derivative of the
free energy
\begin{equation}
\label{deriv}
  \mu(\varepsilon) \equiv
  \frac{\delta F[f(\varepsilon)]}{\delta f(\varepsilon)}.
\end{equation}
From Eq.~\ref{split}, $\mu(\varepsilon)$ can be written in two parts
\begin{equation}
\label{splitmu}
  \mu(\varepsilon) = \mu^{{\rm id}}(\varepsilon) + \mu^{{\rm ex}}(\varepsilon).
\end{equation}
Functional differentiation of Eq.~\ref{ideal} yields
\begin{equation}
\label{muideal}
  \mu^{{\rm id}}(\varepsilon) = \ln \left[ \frac{f(\varepsilon)}{V} \right].
\end{equation}
Collecting together Eqs.~\ref{equilib}, \ref{splitmu} and \ref{muideal}
gives the ratios of densities in any two of the ${\cal M}$ coexisting
phases
\begin{equation}
\label{ratio}
  \frac{f(\varepsilon)_{{\cal B}}/V_{{\cal B}}}
  {f(\varepsilon)_{{\cal A}}/V_{{\cal A}}} =
  \exp \left( \mu^{{\rm ex}}(\varepsilon)_{{\cal A}}
  \raisebox{0pt}[10pt]{$-$} \mu^{{\rm ex}}(\varepsilon)_{{\cal B}} \right)
\end{equation}
in terms of the excess parts of the chemical potentials. Thus, all but one
distribution can be eliminated from Eq.~\ref{conserve}, yielding the
solution for any given phase
\begin{mathletters}
  \begin{equation}
  \label{sim1}
	f(\varepsilon)_{{\cal A}} = f_{P}(\varepsilon) \left/
	\sum_{{\cal B}=1}^{{\cal M}} \frac{V_{{\cal B}}}{V_{{\cal A}}}
	\exp \left( \mu^{{\rm ex}}(\varepsilon)_{{\cal A}}
	\raisebox{0pt}[10pt]{$-$} \mu^{{\rm ex}}(\varepsilon)_{{\cal B}}
	\right) \right.
  \end{equation}
  \begin{equation}
  \label{sim2}
	\mbox{where }\hspace{6mm}
	\mu^{{\rm ex}}(\varepsilon) \equiv
	\frac{\delta F^{{\rm ex}}[f(\varepsilon)]}{\delta f(\varepsilon)}.
	\hspace{25mm}
  \end{equation}
Given a knowledge of $F^{{\rm ex}}$, which specifies the interactions in
the system, (and of the phase volumes), Eqs.~\ref{sim1} and \ref{sim2}
represent a complete solution to the problem. However, they constitute
an uncountable infinity of non-linear simultaneous equations. This is
the source of the mathematical complexity mentioned earlier.
\end{mathletters}

Some simplification is achieved by making a change of variables. Rather
than expressing a thermodynamic state in terms of the densities of the
individual species of particles, $f(\varepsilon)/V$, let us use {\em
moments} of this distribution (as in
\cite{Sollich98,Warren98,Barrat86}). The thermodynamic variables
\begin{equation}
\label{moment}
  \rho_{\alpha} \equiv \int_{-\infty}^{\infty} \varepsilon^{\alpha} \,
	\frac{f(\varepsilon)}{V} \, d\varepsilon \; ;
	\;\;\;\;\;\;\; \alpha=0,1,...,\infty
\end{equation}
will be called `moment densities'. Note that
$\rho_{\alpha}=\overline{\varepsilon^{\alpha}}\rho$, so that $\rho_{0}$
is the overall number density $\rho$. [Mean powers of the size
parameter, $\overline{\varepsilon^{\alpha}}$, are moments of the {\em
normalized} distribution $p(\varepsilon)\equiv f(\varepsilon)/N$.] Each
moment density, being a linear combination of conserved species
densities, is itself conserved and, accordingly, respects the usual
equilibrium conditions. For instance, each `moment chemical potential',
defined by $\mu_{\alpha}\equiv\partial (F/V)/\partial\rho_{\alpha}$, is
equal in coexisting phases. This is clear from expanding the species
chemical potential in partial derivatives
\begin{equation}
\label{powerseries}
  \mu(\varepsilon)\equiv\frac{\delta F}{\delta f(\varepsilon)}
   = \sum_{\alpha=0}^{\infty} \frac{\partial F}{\partial\rho_{\alpha}}
   \frac{\delta\rho_{\alpha}}{\delta f(\varepsilon)}
   = \sum_{\alpha=0}^{\infty}\mu_{\alpha}\,\varepsilon^{\alpha}\;.
\end{equation}
Thus, equality of $\mu(\varepsilon)$ in coexisting phases requires
equality of each $\mu_{\alpha}$.

We now have a discrete set of thermodynamic variables, and can
substitute the power series expression (Eq.~\ref{powerseries}) for
$\mu(\varepsilon)$ into Eq.~\ref{sim1}, yielding
\begin{mathletters}
  \begin{equation}
  \label{sim3}
	f(\varepsilon)_{{\cal A}} = f_{P}(\varepsilon) \left/
	\sum_{{\cal B}=1}^{{\cal M}} \frac{V_{{\cal B}}}{V_{{\cal A}}}
	\exp \left( \sum_{\alpha=0}^{\infty}
	(\mu^{{\rm ex}}_{\alpha{\cal A}}-\mu^{{\rm ex}}_{\alpha{\cal B}})
	\,\varepsilon^{\alpha} \right) \right.
  \end{equation}
  \begin{equation}
  \label{sim4}
	\mbox{with }\hspace{16mm}
	\mu^{{\rm ex}}_{\alpha} \equiv
	\frac{\partial F^{{\rm ex}}/V}{\partial\rho_{\alpha}}
	\hspace{25mm}
  \end{equation}
which, with Eq.~\ref{moment}, form a countable infinity of simultaneous
equations. The excess free energy is now a {\em function} $F^{{\rm
ex}}(\rho_{0},\rho_{1},...)$ of the moment densities.
\end{mathletters}

The equations thus far are perfectly general, but the advantages of this
formalism become apparent when we consider a narrow distribution of
sizes, {\it i.e}.~a system which is close to monodisperse. If the origin
for the parameter $\varepsilon$ is chosen (by fixing the reference
$R_{0}$) to be close to the centre of the narrow distribution, then
$\varepsilon$ is a {\em small} number for most if not all particles.
Hence in Eq.~\ref{sim3}, $f_{P}(\varepsilon)$ vanishes for large
$\varepsilon$, so the power series in the denominator becomes a
well-controlled expansion. The results have a particularly simple form if
the origin is chosen to be the mean of the parent distribution, so that
$\overline{\varepsilon}_{P} \equiv 0$. Henceforth this choice is
assumed.

The solution to Eqs.~\ref{moment}, \ref{sim3} and \ref{sim4} is now
calculated to first order in $\varepsilon$. This will yield the exact
phase equilibria in the limit of a narrow parent,
$\overline{\varepsilon^2}_{P}\to 0$. Expanding Eq.~\ref{sim3} to first
order and integrating over $\varepsilon$ gives
\begin{displaymath}
  \sum_{{\cal B}=1}^{{\cal M}} \frac{V_{{\cal B}}}{V_{{\cal A}}}
	\exp(\mu_{0{\cal A}}^{{\rm ex}}-\mu_{0{\cal B}}^{{\rm ex}})
   	= \frac{N}{N_{{\cal A}}} \left[ 1 \raisebox{0pt}[10pt]{$+$}
	O(\varepsilon^2) \right].
\end{displaymath}
To zeroth order, Eq.~\ref{ratio} gives
\begin{displaymath}
  \frac{V_{{\cal B}}}{V_{{\cal A}}}
	\exp(\mu_{0{\cal A}}^{{\rm ex}}-\mu_{0{\cal B}}^{{\rm ex}})
   	= \frac{N_{{\cal B}}}{N_{{\cal A}}} \left[ 1
	\raisebox{0pt}[10pt]{$+$} O(\varepsilon) \right].
\end{displaymath}
Note the different orders of expansion. Substituting these expressions
back into \ref{sim3} yields
\begin{equation}
\label{each}
  \frac{f(\varepsilon)_{{\cal A}}}{N_{{\cal A}}}
	= \frac{f_{P}(\varepsilon)}{N} \,
	\left[ 1-\varepsilon\,\mu_{1{\cal A}}^{{\rm ex}}
	+ \frac{\varepsilon}{N} \sum_{{\cal B}=1}^{{\cal M}}
	N_{{\cal B}}\,\mu_{1{\cal B}}^{{\rm ex}}
	+ O(\varepsilon^2) \right].
\end{equation}
To obtain Eq.~\ref{each}, the {\em prefactor} of $f_{P}(\varepsilon)$
in Eq.~\ref{sim3} is expanded to first order in $\varepsilon$, but the
distribution $f_{P}(\varepsilon)$ itself remains exact.
Thus, other than narrowness, no limitations are put on the form of
$f_{P}(\varepsilon)$. Any distribution can be treated, however
asymmetric or discontinuous. The population may even contain
finite amounts of some components, contributing delta functions to
$f_{P}(\varepsilon)$.

In Eq.~\ref{each} we see that the distribution in any given phase ${\cal A}$
depends, as one would expect, on the properties of all the other ${\cal M}$
phases with which it coexists. However, taking the difference
(denoted $\Delta$) between the {\em normalized} distributions in
{\em any} two of the ${\cal M}$ coexisting phases, we find the
strikingly simple expression
\begin{equation}
\label{limit}
  \Delta p(\varepsilon) \to -\varepsilon \,
	p_{\raisebox{-3pt}{\scriptsize$P$}}(\varepsilon) \:
	\Delta\mu_{1}^{{\rm ex}}
\end{equation}
in the limit as $\overline{\varepsilon^2}_{P}\to 0$, where
$p_{\raisebox{-3pt}{\scriptsize$P$}}(\varepsilon)$ is the normalized parent
distribution. [Note that the solution for each
phase (Eq.~\ref{each}) is recoverable from the neater sum
(Eq.~\ref{conserve}) and difference (Eq.~\ref{limit}) equations.]
Surprisingly we have found that, in the multi-phase system, the
difference in compositions of any pair of phases
is identical to the expression found earlier \cite{Evans98} for
two-phase coexistence. Thus the same universal laws follow
\cite{Evans98}, relating any pair of phases. This is {\em not} an
obvious result, since the parent appearing in Eq.~\ref{limit} is the
combined population of the whole system, not just of the two phases in
question as it is in the two-phase coexistence problem.

Equation \ref{limit} is very generally applicable. It is valid
for any system with a narrow distribution (that is, narrower than the
range of linearizability of the fugacity), whatever particles or
interactions it comprises. Furthermore, recall that $\varepsilon$ need
not parameterize size deviations, but could represent charge, mass or
any other sole polydisperse quantity. By analysing multi-phase
coexistence, we have found that Eq.~\ref{limit} does not even depend on
${\cal M}$, the number of phases present.

We have considered a system in which a slightly polydisperse fluid
component is partitioned among several phases. The coexistence of more than
two phases may be the result of tuning the temperature to the triple point
of the monodisperse reference system. Alternatively, the slightly
polydisperse particles may be in the presence of other, dissimilar
components which, by the Gibbs phase rule (which states that an
$n$-component mixture can exhibit up to $n+1$ coexisting phases at
arbitrary temperature), can induce multi-phase coexistence
\cite{widthcomment}. Within such a
multi-component system, a particular, slightly polydisperse component will
respect the above relations, which may be tested by an experimental probe
which is `blind' to the other components. For instance, near-monodisperse
colloidal particles in the presence of `depletant' species \cite{depletion}
exhibit multiple phases. Light scattered only from the near-monodisperse
colloid contains information on its fractionation \cite{scatter},
which should obey
Eq.~\ref{limit}. As an illustration, a multi-phase colloidal sample with
the composition shown in the figure, will obey the above relations, applied
only to those particles in range $X$, with the origin of $\varepsilon$
defined at its centre. The relations are equally applicable to particles in
range $Y$, if we are blind to all other particles ({\it e.g}.~they may be
made invisible by matching their refractive index to that of the solvent),
and redefine $\varepsilon=0$ appropriately.

The form of the solution in Eq.~\ref{limit} is of interest in itself, not
least for the non-appearance of ${\cal M}$. However, one quantity
remains unknown: the constant of proportionality $\Delta\mu_{1}^{\rm ex}$.
That constant is system-dependent. For some substances, $\mu_{1}^{{\rm ex}}$
can be calculated using thermodynamic perturbation theory
\cite{Evans98,Evans99}. Unfortunately, this is not possible for a system of
hard-spheres, as it's Hamiltonian is non-differentiable. Since the hard-sphere
system is of great practical interest for modelling systems with repulsive
interactions, the constant of proportionality is now calculated for that case.

\begin{figure}[h]
  \displaywidth\columnwidth
  \epsfxsize=7.5cm
  \begin{center}
  \leavevmode\epsffile{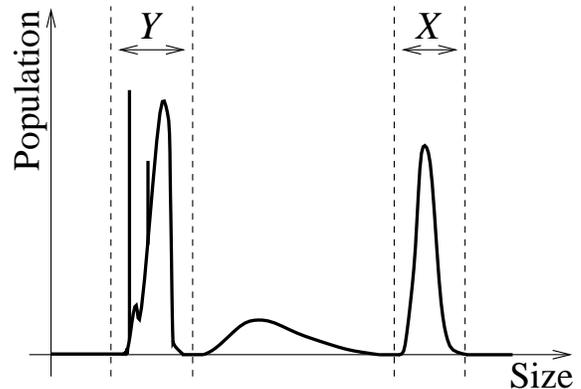}
  \caption{The composition of a sample which could exhibit multi-phase
	coexistence. The narrow part of the distribution in range $X$
	can be treated by the present theory, as could the part in range $Y$.}
  \label{figure}
  \end{center}
\end{figure}

The excess part of the free energy of the polydisperse hard-sphere fluid can
be Taylor expanded in the small size parameter of each of the $N$ particles
of interest thus
\begin{displaymath}
  F^{\rm ex} = F^{\rm ex}_{\rm mono} + \sum_{i=1}^{N} \varepsilon_{i} \,
	\left. \frac{\partial F^{\rm ex}}{\partial \varepsilon_{i}}
	\right|_{\varepsilon_{i}=0} + O(\varepsilon^2)
\end{displaymath}
where $F^{\rm ex}_{\rm mono}$ is the excess free energy of the reference
component of monodisperse hard spheres (in the presence of the rest of the
system --- see Fig.~\ref{figure}). In the reference component, all particles
are alike, so the differentiation may be performed on particle number 1 only,
without loss of generality, giving
\begin{displaymath}
  F^{\rm ex} = F^{\rm ex}_{\rm mono} + N \overline{\varepsilon} \,
	\left. \frac{\partial F^{\rm ex}}{\partial \varepsilon_{1}}
	\right|_{\varepsilon_{1}=0} + O(\varepsilon^2).
\end{displaymath}
The change in the identity (the species) of particle 1 when its size is
varied affects only $F^{\rm id}$. The excess free
energy contains the physical effect of the particle's size on the rest of the
system. By its presence in the container, particle 1 simply excludes other
particles from a volume $V_{\rm excl}$, given that its interactions are purely
hard and repulsive. Thus, increasing its size reduces the effective system
volume, so
\begin{equation}
\label{dFde}
  \frac{\partial F^{\rm ex}}{\partial \varepsilon_{1}} =
	-\frac{\partial F^{\rm ex}}{\partial V} \,
	\frac{d V_{\rm excl}}{d \varepsilon_{1}}.
\end{equation}
In fact the volume from which particle 1 excludes other particles,
$V_{\rm excl}$, depends on their species, so the quantity in Eq.~\ref{dFde}
is a net effective value, {\em defined} by the equation. For the special
case of an almost pure hard sphere system ({\em not} in the presence of
other, dissimilar components),
$V_{\rm excl}=\frac{4}{3}\pi \overline{R}_{P}^3 (2+\varepsilon_{1})^3$ at low
density (correct up to second virial coefficient). At high density, the
geometry of high-order inter-particle interactions modifies this.
In any case, $dV_{\rm excl}/d\varepsilon_{1}$ is of order
a particle volume. The resulting excess free energy density of a polydisperse
hard sphere fluid is
\begin{equation}
  \frac{F^{\rm ex}}{V} = \frac{F^{\rm ex}_{\rm mono}}{V}
	+ 12 \rho_{1} P^{\rm ex} V^{\rm eff} + O(\varepsilon^2)
\end{equation}
where $V^{\rm eff}$ is some (unknown) effective volume, of order
the volume of an average sphere, and exactly that for a near-pure,
low-density system. Applying Eq.\ \ref{sim4} yields
\begin{equation}
  \mu_{1}^{\rm ex} = 12 \, P^{\rm ex} \, V^{\rm eff}
\end{equation}
in terms of the system's excess pressure $P^{\rm ex}$ over an ideal gas.
Since coexisting phases have the same total pressure, it follows that
$\Delta\mu_{1}^{\rm ex}=-12 \, V^{\rm eff} \, \Delta P^{\rm id}$.
So the constant of proportionality in Eq.\ \ref{limit} is
\begin{equation}
  \Delta\mu_{1}^{\rm ex}=-12 \, V^{\rm eff} \, \Delta\rho
\end{equation}
for hard spheres in ergodic (fluid) phases. This calculation contains the
lowest-order effects of polydispersity. Once the polydispersity is sufficient
to alter the mode of packing ({\it e.g}.~small particles preferentially
filling the gaps between big ones), higher-order analysis is needed.

It is apparent
that combining a moment description with a small-variable expansion in
the distribution's width is a productive way to analyse polydisperse
systems. While the applications of this study are clearly wide-ranging,
it is intended to extend its scope by analysing correlation functions
and multiply-polydisperse systems \cite{Evans99}. In addition, some work
is required, using higher-order analysis, to establish the radius of
convergence of the expansion, and quantify more precisely the method's
regime of validity.

{\it Acknowledgements}
Many thanks for informative discussions go to Michael Cates, Peter Sollich,
David Fairhurst, Patrick Warren and Wilson Poon. The work was
funded by the EPSRC (GR/K56025) and a Royal Society of Edinburgh SOEID
Research Fellowship.

\end{multicols}
\end{document}